\documentclass[preprint,showpacs]{revtex4}
\usepackage{amsmath}
\usepackage{graphicx}

\begin{document}

\title{Next-to-leading order QCD predictions for graviton and photon associated production in the Large
Extra Dimensions model at the LHC}
\author{Xiangdong Gao}
\address{Department of Physics
and State Key Laboratory of Nuclear Physics and Technology, Peking
University, Beijing 100871, China}
\author{Chong Sheng Li}
\email{csli@pku.edu.cn}
\address{Department of Physics and State
Key Laboratory of Nuclear Physics and Technology, Peking University,
Beijing 100871, China}
\author{Jun Gao}
\address{Department of Physics and
State Key Laboratory of Nuclear Physics and Technology, Peking
University, Beijing 100871, China}
\author{Robert J. Oakes}
\address{Department of Physics and Astronomy, Northwestern
University, Evanston, Illinois, 60208-3112, USA}
\author{Jian Wang}
\address{Department of Physics
and State Key Laboratory of Nuclear Physics and Technology, Peking
University, Beijing 100871, China}
\date{\today}

\begin{abstract}
We present the calculations of the complete next-to-leading
order(NLO) QCD corrections to the inclusive total cross sections for
the Kaluza-Klein(KK) graviton and photon associated production
process $pp \rightarrow \gamma G_{KK} + X$ in the large extra
dimensions model at the LHC. We show that the NLO QCD
corrections in general enhance the total cross sections and reduce
the dependence of the total cross sections on the factorization and
renormalization scales. When jet veto is considered, the NLO
corrections reduce the total cross sections. We also calculate some
important differential cross sections for this process at NLO: the
missing transverse momentum distribution, the transverse momentum
distribution and the pseudorapidity distribution of photon.
\end{abstract}

\maketitle

\section{introduction}
The idea that the extra dimensions theory can appear at the TeV scale,
well below the Planck scale $M_P \sim 1.2\times10^{19}$GeV, was
proposed in the 1990s\cite{ArkaniHamed:1998rs, ArkaniHamed:1998nn,
Antoniadis:1998ig, Randall:1999ee, Randall:1999vf, Lykken:1996fj,
Witten:1996mz, Horava:1995qa, Horava:1996ma, Antoniadis:1990ew}
and promises rich phenomenology at the TeV scale. Now the search for extra
dimensions is one of the important tasks for the LHC.

Among various extra dimensions models the large extra
dimensions(LED) model introduced in Ref.\cite{ArkaniHamed:1998rs,
ArkaniHamed:1998nn, Antoniadis:1998ig} is the first TeV scale
gravity theory and has been extensively studied\cite{Han:1998sg,
Giudice:1998ck}. In this model space-time has $4+\delta$ dimensions
and the standard model(SM) particles reside in the usual
$3+1$-dimensional SM brane and can not propagate in the extra
$\delta$-dimensional space, which is assumed to be compacted on a
torus with a common radius $R$, while gravity can propagate in the
whole $4+\delta$ dimensional world. From the view of our 4-dimensional world,
there exists infinitely many Kaluza-Klein(KK) modes of
gravitons with mass $|k|/R$ interacting with SM particles, where
$k^2=\sum_{i=1}^{\delta} k_i^2$ with $k_i$ being the integer. The
4-dimensional Planck scale is no longer the fundamental scale, but
an effective scale in the 4-dimensional world, and is related to a
fundamental scale $M_D \sim$TeV by the Newtonian law of gravitation
in $4+\delta$ dimensions\cite{ArkaniHamed:1998rs, Giudice:1998ck}
\begin{equation}
\bar{M}_P \equiv M_P/\sqrt{8\pi} =
R^{\delta/2}M_D^{1+\delta/2},\label{scalerelation}
\end{equation}
where $\bar{M}_P$ is the reduced Planck mass. According to
Eq.(\ref{scalerelation}), deviations from the usual Newtonian
gravitational law appear at $R \sim 10^{\frac{32}{\delta}-19}$
meters, which is not a conflict with the current gravitational
experiments\cite{Adelberger:2002ic} once $\delta \geq 2$. Before
further results of terrestrial gravitational experiments appear, one
may resort to colliders to find signals of this model. Although, in
this model the couplings of gravitons to SM particles are suppressed
by $1/M_P$ \cite{Han:1998sg, Giudice:1998ck}, the summation of the
production of large numbers of KK modes with arbitrary mass smaller
than $M_D$ may compensate for this suppression and lead to
observable effects. There are two ways to probe such effects at
colliders: graviton emission and virtual graviton exchange, which
have been investigated in Ref. \cite{Giudice:1998ck, Han:1998sg,
Cheung:2004ab}. As shown in Ref.\cite{Giudice:1998ck}, because of
the suppression of the couplings of gravitons to SM particles, the
decays of gravitons to SM particles have a small probability to
occur before they propagate into the extra $\delta$-dimensional
space, which means the gravitons behave like massive, stable, and
noninteracting particles once they are produced. Thus, the signal
for graviton and  photon associated production at the LHC is a
single photon plus missing energy. Since the electromagnetic
coupling is small and the $q\bar{q}$ luminosity is lower than for
$gg$ at the LHC the rate for this process is much smaller than for
jet and graviton associated production. But the photon signal would
be a clean signature; and in case of discovery in jet plus graviton
events, the photon plus graviton signal would provide a useful
independent test\cite{Giudice:1998ck}. Only leading-order(LO)
calculations and analysis of the process were performed in
Ref.\cite{Giudice:1998ck}. Since LO cross sections for processes at
hadron colliders suffer from large uncertainties due to the choices
of the renormalization scale ($\mu_r$) and factorization scale
($\mu_f$) and higher order QCD corrections are generally large and
can improve the scale uncertainties at hadron colliders. Several
works\cite{Mathews:2004xp, Mathews:2005bw, Mathews:2005zs,
Kumar:2006id, Li:2006yv, Kumar:2008pk, Kumar:2009nn, Agarwal:2009zg,
Karg:2009xk} have performed next-to-leading order(NLO) QCD corrections in extra dimensions
models. In this paper we present the complete calculations of NLO
QCD corrections to this process which improve the theoretical
predictions.

This paper is organized as follows. In Sec.\ref{capt:tree} we show
the analytic results of the LO calculations and define the
notation. In Sec.\ref{capt:nlo} we present the details of the
calculations of both the virtual and real parts of the NLO QCD
corrections. In Sec.\ref{capt:nu} we give the numerical predictions
for inclusive and differential cross sections at the LHC. We close
this paper with a brief conclusion. For completeness, the relevant
Feynman rules are collected in Appendix A and the lengthy analytic
expressions of the results of our calculations are summarized in
Appendix B.
\section{Leading-order calculations}\label{capt:tree}
The KK gravitons with different masses can be produced at colliders only
if kinematically allowed. Contributions from the different KK modes then
must be summed up. Since the KK graviton mass separation of
$\mathcal{O}(1/R)$ is much smaller than all the other physical
scales involved, we can replace the discrete summation of different KK modes
by a continuous integration. In general, the differential cross
section for graviton production can be expressed
as\cite{Giudice:1998ck}
\begin{equation}
\frac{d^2\sigma}{dt~dm} = S_{\delta - 1}
\frac{\bar{M}^2_P}{M^{2+\delta}_D} m^{\delta - 1}
\frac{d\sigma_m}{dt}
\end{equation}
with
\begin{equation}
S_{\delta-1} = \frac{2\pi^{\delta/2}}{\Gamma(\delta/2)},
\end{equation}
where $S_{\delta-1}$ is the surface of a unit-radius sphere in
$\delta$ dimensions and $d\sigma_{m}/dt$ is the differential cross
section for producing a single KK graviton of mass $m$. Throughout this work
we perform the integration on $m$ from $0.1M_D$ to $M_D$.

\begin{figure}
\includegraphics[scale=0.35]{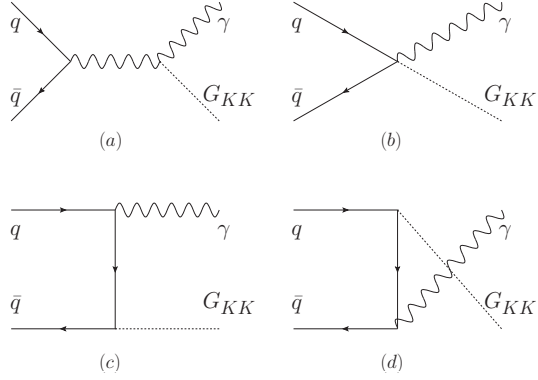}
\caption{\label{tree}Leading-order Feynman diagrams for $q\bar{q}
\rightarrow \gamma G_{KK}$.}
\end{figure}

The leading-order Feynman diagrams for the graviton and photon
associated production process  $q(p_1) \bar{q}(p_2) \rightarrow
\gamma(p_3) G_{KK}(p_4)$ are shown in Fig. \ref{tree}. The related
Feynman rules are given in Ref. \cite{Han:1998sg, Giudice:1998ck}
and are collected in Appendix A. The LO amplitudes have been given
in Ref.\cite{Han:1998sg} so we need only to show the amplitudes squared
here. In the LED model the spin sum over the polarization tensors of
the graviton is
\begin{equation}
\sum^5_{s=1} \epsilon^s_{\mu \nu} \epsilon^{s\ast}_{\alpha \beta} =
P_{\mu \nu \alpha \beta},
\end{equation}
with
\begin{equation}
P_{\mu \nu \alpha \beta} = \eta_{\mu \alpha}\eta_{\nu \beta} +
\eta_{\mu \beta}\eta_{\mu \alpha} - \frac{2}{n-1} \eta_{\mu
\nu}\eta_{\alpha \beta} + ...,
\end{equation}
where the dots represent terms proportional to the graviton momentum
$p^{\mu}_4$ and  $p^{\mu}_4 T_{\mu \nu} = 0$, giving no
contribution to the amplitude. We performed our calculations in $n =
4 - 2\epsilon$ dimensions.

The LO partonic cross section is then
\begin{equation}
\hat{\sigma}_m^{B} = \frac{1}{2s}\int d\Gamma_2
\overline{\sum}|M^B|^2,
\end{equation}
with
\begin{eqnarray}
\overline{\sum}|M^B|^2 & = & \frac{\text{eQ}^2 \kappa ^2  }{24 s t
u}\left[(s+4 t) m^6-6 t (s+2 t) m^4+\left(s^3+6 t s^2+18 t^2 s+16
t^3\right) m^2 \right.\nonumber \\ && \left. -4 t \left(s^3+3 t
s^2+4 t^2 s+2 t^3\right)\right],\label{Born2}
\end{eqnarray}
where $\kappa = \sqrt{2}/\bar{M}_{P}$, $s$, $t$, and $u$ are the
Mandelstam variables defined as
\begin{equation}
s = (p_1 + p_2)^2,~~~t = (p_1-p_3)^2,~~~u = (p_1 - p_4)^2,
\end{equation}
and $\overline{\sum}$ means that the colors and spins of the
outgoing particles have been summed over and the colors and spins of
the incoming particles have been averaged over. Equation(\ref{Born2}) is a
coincidence with the result shown in Ref.\cite{Giudice:1998ck}.

The LO total cross section can be obtained by convoluting the
partonic cross sections with the parton distribution functions(PDF)
$G_{q,\bar{q}}$ in the protons:
\begin{equation}
\sigma^{B}_m = \int dx_1 dx_2 [G_{q/p}(x_1,\mu_{f})
G_{\bar{q}/p}(x_2,\mu_{f}) + G_{q/p}(x_2,\mu_{f})
G_{\bar{q}/p}(x_1,\mu_{f})] \hat{\sigma}_m^{B}.
\end{equation}
Here $\mu_f$ is the factorization scale.

\section{Next-to-leading order calculations}
\label{capt:nlo}The NLO corrections to the associated production of
a graviton and a photon can be separated into the virtual
corrections arising from loop diagrams of colored particles and the
real corrections arising from the radiation of a real gluon or a
massless (anti)quark. We carried out the calculations in the 't
Hooft-Feynman gauge and used dimensional
regularization\cite{'tHooft:1972fi} in $n = 4-2\epsilon$ dimensions
to regulate all the ultraviolet(UV), soft and collinear divergences.
We performed two independent calculations for both the analytical
and numerical results for cross checking, and the results of
the two groups agree with each other.

\subsection{Virtual corrections}
\begin{figure}
\scalebox{0.35} {\includegraphics*[10,-300][1200,900]{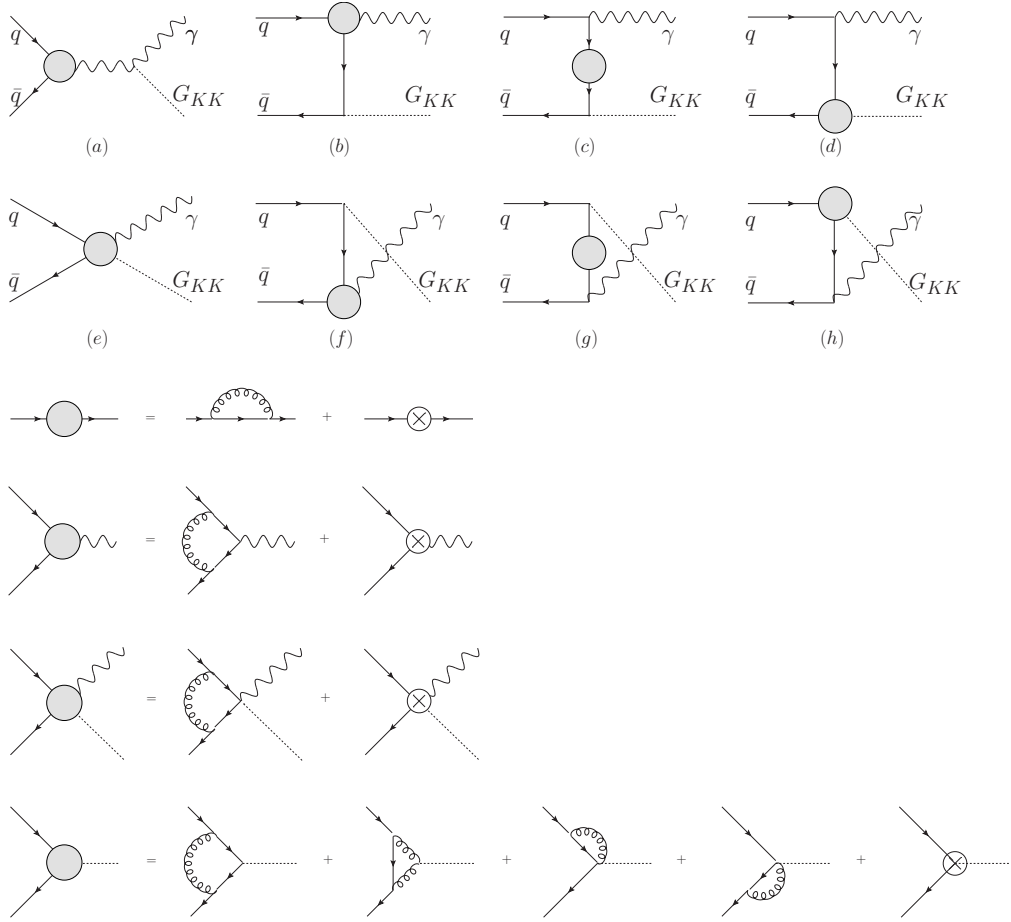}}
\caption{\label{loop}One-loop virtual diagrams including vertex and
self-energy corrections to $q\bar{q} \rightarrow \gamma G_{KK}$.
Each brown vertex is UV divergence free.}
\end{figure}

\begin{figure}
\scalebox{0.4}
{\includegraphics*[10,300][1000,800]{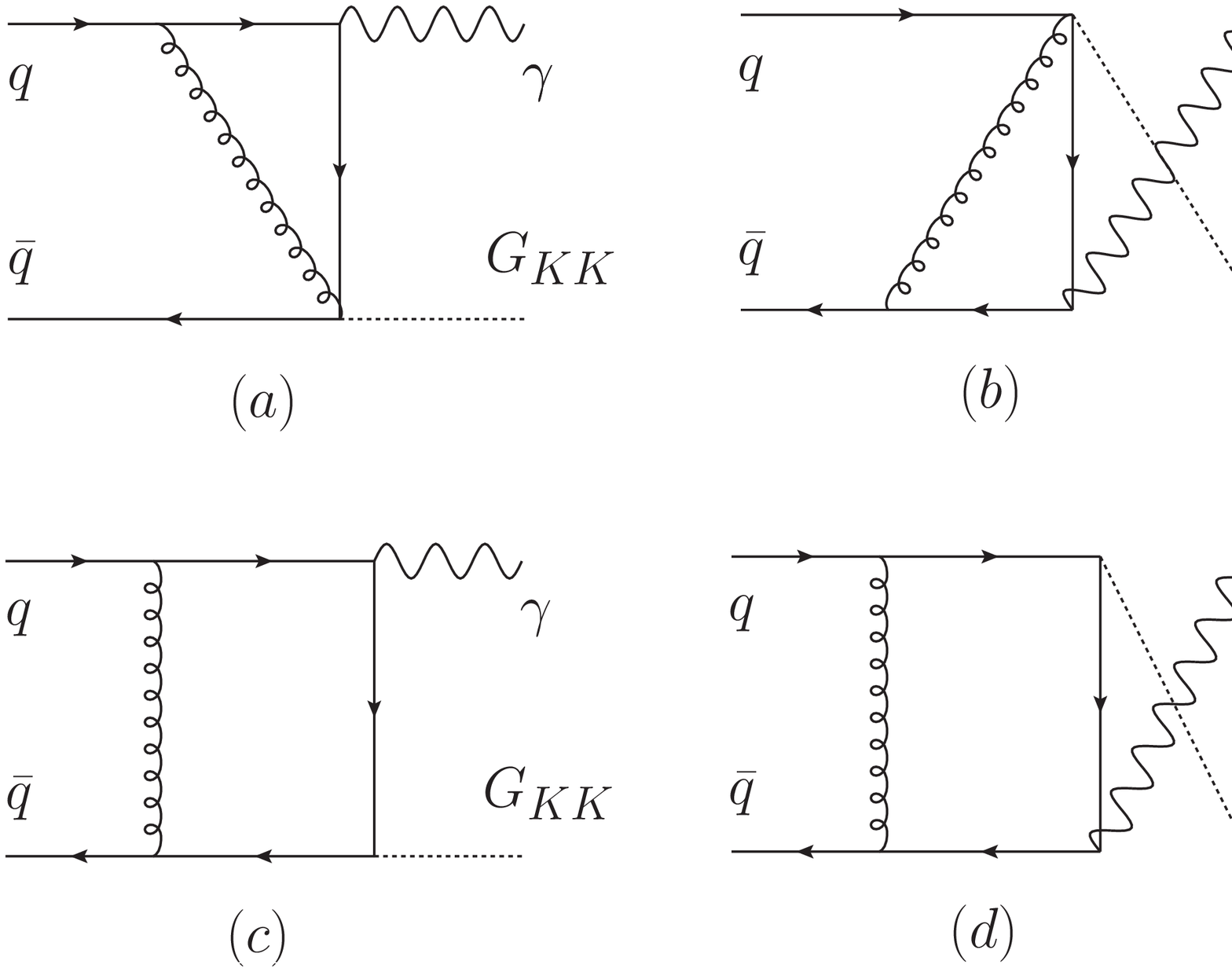}}
\caption{\label{trianglebox}Box and triangle diagrams for $q\bar{q}
\rightarrow \gamma G_{KK}$. The UV divergences cancel among
the five diagrams.}
\end{figure}

The Feynman diagrams for the virtual corrections to $q\bar{q}
\rightarrow \gamma G_{KK}$ are shown in Fig.\ref{loop} and
Fig.\ref{trianglebox}. They consist of self-energy, vertex, triangle
and box diagrams. In order to remove the UV divergence we adopt the
on-shell renormalization scheme\cite{Sirlin:1980nh, Marciano:1980pb,
Sirlin:1981yz, Aoki:1982ed}.

We denote the bare and renormalized quark wave functions by
$\psi_{q0}$ and $\psi_q$, respectively. The renormalization constant
$\delta Z_q$ is then defined by
\begin{equation}
\psi_{q0} = (1+\delta Z_q)^{1/2} \psi_q.
\end{equation}
Calculating the quark self-energy diagram we obtain the explicit
expression for $\delta Z_q$:
\begin{equation}
\delta Z_q = \frac{\alpha_s}{4\pi} C_F (\frac{1}{\epsilon} -
\frac{1}{\epsilon_{UV}}).
\end{equation}
Here $C_F = \frac{4}{3}$ while $1/\epsilon$ and $1/\epsilon_{UV}$
represent infrared(IR) and UV divergences, respectively.

After renormalization the UV divergences in the virtual
corrections are removed leaving the IR divergences and the finite
terms. The $\mathcal{O} (\alpha_s)$ virtual corrections to the
partonic total cross section can then be expressed as
\begin{equation}
\label{eq:virtual} \hat{\sigma}^V_m  =  \frac{1}{2s}\int d\Gamma_2
\overline{\sum}\left[
\frac{\alpha_s}{2\pi}\frac{\Gamma(1-\epsilon)}{\Gamma(1-2\epsilon)}
\left(\frac{4\pi\mu^2_r}{s}\right)^{\epsilon}\left(\frac{2A^V_2}{\epsilon^2}+\frac{2A^V_1}{\epsilon}\right)
|M^B|^2 + \frac{\alpha_s}{2\pi}\mathcal{S}\right],
\end{equation}
with
\begin{equation}
A^V_2 = -C_F,~~~A^V_1 = -\frac{3}{2}C_F.
\end{equation}
Here $\mathcal{S}$ represents finite terms in the virtual corrections
and the explicit expressions are given in Appendix B. The
cancellation of IR divergent terms $1/{\epsilon^2}$ and
$1/\epsilon$ will be discussed in detail below.

\subsection{Real gluon emission}
\begin{figure}
\scalebox{0.5} {\includegraphics*[10,250][700,800]{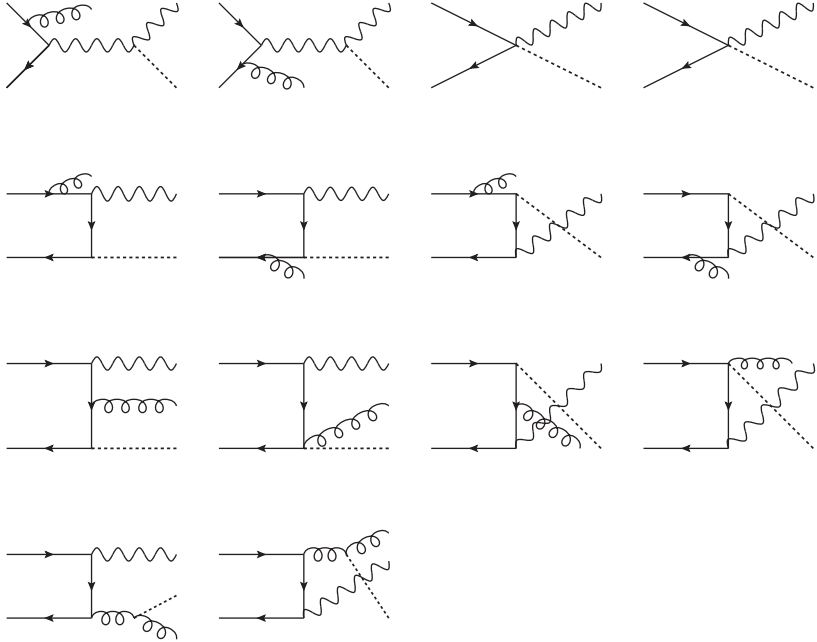}}
\caption{\label{realgluon}Feynman diagrams for $q\bar{q} \rightarrow
\gamma G_{KK} + g$.}
\end{figure}

The Feynman diagrams for the real gluon emission process $q(p_1)
\bar{q}(p_2) \rightarrow \gamma(p_3) G_{KK}(p_4) + g(p_5)$ are shown
in Fig. \ref{realgluon}.

The phase space integration for the real gluon emission will produce
soft and collinear singularities which can be conventionally
isolated by slicing the phase space into different regions using
suitable cutoffs. In this paper we use the two-cutoff phase space
slicing method\cite{Harris:2001sx}, which introduces two arbitrary
small cutoffs; i.e., a soft cutoff $\delta_s$ and a collinear cutoff
$\delta_c$, to divide the three-body phase space into three
regions.

First, the phase space is separated into two regions by the soft
cutoff $\delta_s$, according to whether the energy of the emitted
gluon is soft; i.e., $E_5\leq \delta_s\sqrt{s}/2$, or hard; i.e.,
$E_5> \delta_s\sqrt{s}/2$. Then the parton level real
cross section $\hat{\sigma}^{R}_{m}$ can be written as
\begin{eqnarray}
\hat{\sigma}^{R}_{m}= \hat{\sigma}^{S}_{m} +\hat{\sigma}^{H}_{m},
\end{eqnarray}
where $\hat{\sigma}^{S}_{m}$ and $\hat{\sigma}^{H}_{m}$ are the
contributions from the soft and hard regions, respectively.
$\hat{\sigma}^{S}_{m}$ contains all the soft divergences, which can
explicitly be obtained after the integration over the phase space of
the emitted gluon. Next, in order to isolate the remaining collinear
divergences from $\hat{\sigma}^{H}_{m}$, the collinear cutoff
$\delta_c$ is introduced to further split the hard gluon phase space
into two regions, according to whether the Mandelstam variables
$t_{i5} \equiv (p_i-p_5)^2$, with $i=1,2$, satisfy the collinear
condition $-\delta_c s< t_{i5}< 0$ or not. We then have
\begin{eqnarray}
\hat{\sigma}^{H}_{m}= \hat{\sigma}^{HC}_{m}+
\hat{\sigma}^{\overline{HC}}_{m},
\end{eqnarray}
where the hard collinear part $\hat{\sigma}^{HC}_{m}$ contains the
collinear divergences, which also can explicitly be obtained after
the integration over the phase space of the emitted gluon. The
hard noncollinear part $\hat{\sigma}^{\overline{HC}}_{m}$ is finite
and can be numerically computed using standard Monte Carlo
integration techniques\cite{Lepage:1977sw} and can be written in
the form
\begin{eqnarray}
d\hat{\sigma}^{\overline{HC}}_{m}=\frac{1}{2s}
\overline{\sum}|M^{q\bar{q}}|^2 d\overline{\Gamma}_3. \label{nonHC}
\end{eqnarray}
Here $d\overline{\Gamma}_3$ is the hard noncollinear region of the
three-body phase space.

In the next two subsections we will discuss in greater detail the soft and
hard collinear gluon emission.

\subsubsection{Soft gluon emission}
In the limit that the energy of the emitted gluon becomes small,
i.e. $E_5\leq \delta_s\sqrt{s}/2$, the amplitude squared
$\overline{\sum}|M(q\bar{q} \to \gamma G_{KK} +g)|^2$ can be
factorized into the Born amplitude squared times an eikonal factor
$\Phi_{\text{eik}}$:

\begin{equation}
\overline{\sum}|M(q\bar{q} \to \gamma G_{KK} +g)|^2
\stackrel{\text{soft}}{\longrightarrow}
(4\pi\alpha_s\mu_r^{2\epsilon}) \overline{\sum}|M^B|^2
\Phi_{\text{eik}},
\end{equation}
where the eikonal factor is given by
\begin{equation}
\Phi_{\text{eik}} = C_F \frac{s}{(p_1 \cdot p_5)(p_2 \cdot p_5)}.
\end{equation}
Moreover, the three-body phase space in the soft limit can also be
factorized:
\begin{equation}
d\Gamma_3(q\bar{q} \to \gamma G_{KK}
+g)\stackrel{\text{soft}}{\longrightarrow}d\Gamma_2(q\bar{q} \to
\gamma G_{KK})dS.
\end{equation}
Here $dS$ is the integration over the phase space of the soft gluon
and is given by\cite{Harris:2001sx}
\begin{equation}
dS = \frac{1}{2(2\pi)^{3-2\epsilon}} \int^{\delta_s \sqrt{s}/2}_0
dE_5E_5^{1-2\epsilon}d\Omega_{2-2\epsilon}.
\end{equation}
The parton level cross section in the soft region can then be
expressed as
\begin{equation}
\label{sigma:soft1} \hat{\sigma}^S_m =
(4\pi\alpha_s\mu^{2\epsilon}_r)\int
d\Gamma_2\overline{\sum}|M^B|^2\int dS \Phi_{\text{eik}}.
\end{equation}
Using the approach in Ref.\cite{Harris:2001sx}, after integration
over the soft gluon phase space, Eq.(\ref{sigma:soft1}) becomes
\begin{equation}
\hat{\sigma}^S_m = \hat{\sigma}^B_m
\left[\frac{\alpha_s}{2\pi}\frac{\Gamma(1-\epsilon)}{\Gamma(1-2\epsilon)
}\left(\frac{4\pi\mu^2_r}{s}\right)^{\epsilon}\right]\left(\frac{A^s_2}
{\epsilon^2}+\frac{A^s_1}{\epsilon}+A^s_0\right)
\end{equation}
with
\begin{equation}
A^s_2 = 2C_F,~~~~A^s_1 = -4C_F \log \delta_s,~~~~A^s_0 = 4C_F \log^2
\delta_s.
\end{equation}
\subsubsection{Hard collinear gluon emission}
In the hard collinear region, $E_5> \delta_s\sqrt{s}/2$ and
$-\delta_c s< t_{i5} < 0$, the emitted hard gluon is collinear to
one of the incoming partons. As a consequence of the factorization
theorem\cite{Collins:1985ue, Bodwin:1984hc} the matrix element
squared for $q\bar{q} \rightarrow \gamma G_{KK} +g$ can be
factorized into the product of the Born amplitude squared and the
Altarelli-Parisi splitting function for $q(\bar{q}) \rightarrow
q(\bar{q})g$\cite{Altarelli:1977zs, Ellis:1980wv, Bergmann:1989zy,
Kunszt:1992tn, Mangano:1991jk}; that is,
\begin{equation}
\overline{\sum}|M(q\bar{q}\rightarrow\gamma G_{KK} +
g)|^2\stackrel{\text{collinear}}
{\longrightarrow}(4\pi\alpha_s\mu^{2\epsilon}_r)\overline{\sum}|M^B|^2\left(\frac{-2P_{qq}(z,\epsilon)}{zt_{15}}
+\frac{-2P_{\bar{q}\bar{q}}(z,\epsilon)}{zt_{25}}\right),
\end{equation}
where $z$ denotes the fraction of the momentum of the incoming
parton carried by $q(\bar{q})$ with the emitted gluon taking a
fraction $(1-z)$. $P_{ij}(z,\epsilon)$ are the unregulated
splitting functions in $n=4-2\epsilon$ dimensions for $0<z<1$ which
can be related to the usual Altarelli-Parisi splitting
kernels\cite{Altarelli:1977zs} as follows: $P_{ij}(z,\epsilon) =
P_{ij}(z)+\epsilon P^{\prime}_{ij}(z)$. Explicitly
\begin{eqnarray}
P_{qq}(z) = P_{\bar{q}\bar{q}}(z) = C_F \frac{1+z^2}{1-z} + C_F \frac{3}{2} \delta(1-z),\\
P^{\prime}_{qq}(z) = P^{\prime}_{\bar{q}\bar{q}}(z) = -C_F (1-z) +
C_F \frac{1}{2} \delta(1-z).
\end{eqnarray}
Moreover, the three-body phase space can also be factorized in the collinear limit
and, for example, in the limit $-\delta_c s < t_{15} < 0$ it has the following form\cite{Harris:2001sx}:
\begin{equation}
d\Gamma_3(q\bar{q}\rightarrow\gamma G_{KK} +
g)\stackrel{\text{collinear}}{\longrightarrow}d\Gamma_2(q(\bar{q})
\rightarrow \gamma G_{KK}; s^{\prime} = zs)
\frac{(4\pi)^{\epsilon}}{16\pi^2\Gamma(1-\epsilon)}dzdt_{15}[-(1-z)t_{15}]^{-\epsilon}.
\end{equation}
Here the two-body phase space should be evaluated at a squared
parton-parton energy of $zs$. Thus the three-body cross section in
the hard collinear region is given by\cite{Harris:2001sx}
\begin{eqnarray}
d\sigma^{HC}_m & = & d\hat{\sigma}^B_m
\left[\frac{\alpha_s}{2\pi}\frac{\Gamma(1-\epsilon)}{\Gamma(1-2\epsilon)}
\left(\frac{4\pi\mu^2_r}{s}\right)^{\epsilon}\right](-\frac{1}{\epsilon})
\delta_c^{-\epsilon}\left[P_{qq}(z,\epsilon)G_{q/p}(x_1/z)G_{\bar{q}/p}(x_2)
\right.\nonumber\\&& \left.+
P_{\bar{q}\bar{q}}(z,\epsilon)G_{\bar{q}/p}(x_1/z)G_{q/p}(x_2) +
(x_1 \leftrightarrow x_2)\right]
\frac{dz}{z}\left(\frac{1-z}{z}\right)^{-\epsilon}dx_1dx_2
\end{eqnarray}
where $G_{q(\bar{q})/p}(x)$ is the bare PDF.
\subsection{Massless (anti)quark emission}
\begin{figure}
\scalebox{0.5} {\includegraphics*[10,300][700,800]{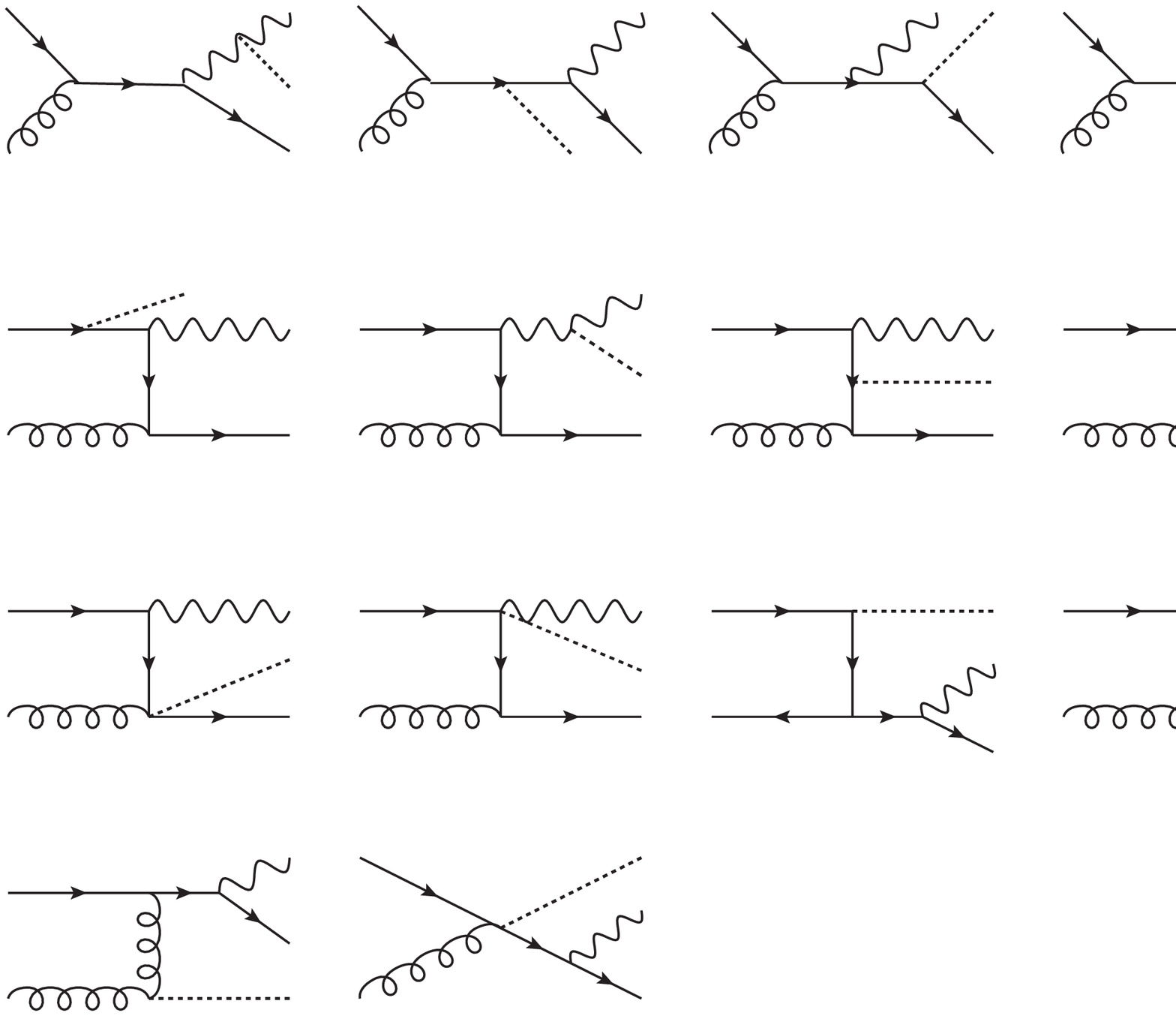}}
\caption{\label{realquark}Feynman diagrams for $qg \rightarrow
\gamma G_{KK} + q$.}
\end{figure}
In addition to real gluon emission a second set of real emission
corrections to the inclusive cross section for $pp\rightarrow \gamma
G_{KK}$ at NLO involves the processes with an additional massless
$q(\bar q)$ in the final state:

\begin{equation}
q(\bar{q})g \rightarrow \gamma G_{KK} + q(\bar{q}).
\end{equation}
The relevant Feynman diagrams are shown in Fig. \ref{realquark}.
The diagrams for $\bar{q}$ emission are similar and are omitted
here.

Since the contributions from real massless $q(\bar{q})$ emission
contain initial state collinear singularities we need to use
the two cutoff phase space slicing method \cite{Harris:2001sx} to
isolate these collinear divergences. But we only split the phase
space into two regions because there are no soft divergences.
Consequently, using the approach in Ref.~\cite{Harris:2001sx}, the
cross sections for the processes with an additional massless
$q(\bar{q})$ in the final state can be expressed as
\begin{eqnarray}
\label{sigma:nc} d\sigma^{add}_m & = &
\sum_{(\alpha=g,\beta=q,\bar{q})} \hat{\sigma}^{\overline{C}}_m
(\alpha \beta \rightarrow \gamma G_{KK} +
q(\bar{q}))[G_{\alpha/p}(x_1)G_{\beta/p}(x_2)+ (x_1 \leftrightarrow
x_2)]dx_1dx_2\nonumber\\ &&+d\hat{\sigma}^B_m
\left[\frac{\alpha_s}{2\pi}\frac{\Gamma(1-\epsilon)}{\Gamma(1-2\epsilon)}
\left(\frac{4\pi\mu^2_r}{s}\right)^{\epsilon}\right]
(-\frac{1}{\epsilon}) \delta_c^{-\epsilon}
\left[P_{qg}(z,\epsilon)G_{g/p}(x_1/z)G_{\bar{q}/p}(x_2) \right. \nonumber\\
&& \left.+ P_{\bar{q}g}(z,\epsilon)G_{q/p}(x_1)G_{g/p}(x_2/z) + (x_1
\leftrightarrow x_2)\right]
\frac{dz}{z}\left(\frac{1-z}{z}\right)^{-\epsilon}dx_1dx_2
\end{eqnarray}
where
\begin{eqnarray}
P_{qg}(z) & = & P_{\bar{q}g}(z) =
\frac{1}{2}[z^2+(1-z)^2],\nonumber\\ P^{\prime}_{qg}(z) & = &
P^{\prime}_{\bar{q}g}(z) = -z(1-z).
\end{eqnarray}
The $\hat{\sigma}^{\overline{C}}_m$ term in Eq. (\ref{sigma:nc})
represents the noncollinear cross sections for the $q(\bar{q})g$
initiated processes which can be written in the form
\begin{equation}
d\hat{\sigma}^{\overline{C}}_m = \frac{1}{2s}
\overline{\sum}|M(q(\bar{q})g
\stackrel{\text{noncollinear}}{\longrightarrow} \gamma G_{KK} +
q(\bar{q}))|^2d\bar{\Gamma}_3,
\end{equation}
where $d\bar{\Gamma}_3$ is the three-body phase space in the
noncollinear region. The other terms in Eq. (\ref{sigma:nc}) are
the collinear singular cross sections.
\subsection{Mass factorization}
After adding the renormalized virtual corrections and the real
corrections, the parton level cross sections still contain collinear
divergences which can be absorbed into a redefinition of the PDFs at
NLO, generally called mass factorization\cite{Altarelli:1979ub,
Collins:1989gx}. This procedure, in practice, means that first we
convolute the partonic cross section with the bare PDF
$G_{\alpha/p}(x)$ and then use the renormalized PDF
$G_{\alpha/p}(x,\mu_f)$ to replace $G_{\alpha/p}(x)$. In the
$\overline{\text{MS}}$ convention the scale-dependent PDF
$G_{\alpha/p}(x,\mu_f)$ is given by \cite{Harris:2001sx}
\begin{eqnarray}
\label{modifiedPDF} G_{\alpha/p}(x,\mu_f) & = & G_{\alpha/p}(x) +
\sum_{\beta}\left(-\frac{1}{\epsilon}\right)\left[
\frac{\alpha_s}{2\pi}\frac{\Gamma(1-\epsilon)}{\Gamma(1-2\epsilon)}
\times \left(\frac{4\pi\mu^2_r}{\mu_f^2}\right)^{\epsilon}\right]\nonumber\\
&& \times \int_x^1 \frac{dz}{z} P_{\alpha\beta}(z) G_{\beta/p}(x/z).
\end{eqnarray}
This replacement will produce a collinear singular counterterm
which is then combined with the hard collinear contributions to give
 Ref.~\cite{Harris:2001sx} the $\mathcal{O}
(\alpha_s)$ expression for the remaining collinear contribution:
\begin{eqnarray}
&& d\sigma^{coll}_m=  d\hat{\sigma}^B_m\bigg[\frac{\alpha_s}{2\pi}
\frac{\Gamma(1-\epsilon)} {\Gamma(1-2\epsilon)}
\bigg(\frac{4\pi\mu^2_r}{s}\bigg)^\epsilon \bigg]
\{\tilde{G}_{q/p}(x_1,\mu_f) G_{\bar{q}/p}(x_2,\mu_f) +
G_{q/p}(x_1,\mu_f) \tilde{G}_{\bar{q}/p}(x_2,\mu_f) \nonumber
\\ && \hspace{1.2cm}
+\sum_{\alpha=q,\bar{q}}\bigg[\frac{A_1^{sc}(\alpha\rightarrow
\alpha g)}{\epsilon} +A_0^{sc}(\alpha\rightarrow \alpha
g)\bigg]G_{q/p}(x_1,\mu_f) G_{\bar{q}/p}(x_2,\mu_f) \nonumber
\\ && \hspace{1.2cm}
+(x_1\leftrightarrow x_2)\} dx_1dx_2,\label{11}
\end{eqnarray}
where
\begin{eqnarray}
&& A_1^{sc}(q\rightarrow qg)=A_1^{sc}(\bar{q}\rightarrow \bar{q}g)=C_F(2\ln\delta_s +3/2), \\
&& A_0^{sc}=A_1^{sc}\ln(\frac{s}{\mu_f^2}), \\
&&
\tilde{G}_{\alpha(=q,\bar{q})/p}(x,\mu_f)=\sum_{\beta=g,\alpha}\int_x^{1-
\delta_s\delta_{\alpha\beta}} \frac{dy}{y}
G_{\beta/p}(x/y,\mu_f)\tilde{P}_{\alpha\beta}(y)
\end{eqnarray}
with
\begin{eqnarray}
\tilde{P}_{\alpha\beta}(y)=P_{\alpha\beta}(y) \ln(\delta_c
\frac{1-y}{y} \frac{s}{\mu_f^2}) -P_{\alpha\beta}'(y).
\end{eqnarray}

Finally, the NLO total cross section for $pp \rightarrow \gamma
G_{KK}$ in the $\overline{\text{MS}}$ factorization scheme is
\begin{equation}
\sigma^{NLO} = \int dm~S_{\delta - 1}
\frac{\bar{M}^2_P}{M^{2+\delta}_D} m^{\delta - 1} \sigma_m^{NLO}
\end{equation}
with
\begin{eqnarray}
&& \sigma^{NLO}_m= \int dx_1dx_2 \{
\bigg[G_{q/p}(x_1,\mu_f)G_{\bar{q}/p}(x_2,\mu_f)+
(x_1\leftrightarrow x_2)\bigg](\hat{\sigma}^{B}_m +
\hat{\sigma}^{V}_m + \hat{\sigma}^{S}_m +
\hat{\sigma}^{\overline{HC}}_m)\} +\sigma^{coll}_m \nonumber
\\ && \hspace{0.4cm} +\sum_{(\alpha=g,\beta=q,\bar{q})}\int dx_1dx_2
\bigg[G_{\alpha/p}(x_1,\mu_f) G_{\beta/p}(x_2,\mu_f)
+(x_1\leftrightarrow x_2)\bigg]
\hat{\sigma}^{\overline{C}}_m(\alpha\beta\rightarrow \gamma G_{KK}
+ \beta) \, .\label{totalXsec}
\end{eqnarray}
Note that the above expression contains no singularities since
$2A_2^V + A_2^s = 0$ and $2A_1^V + A_1^s + A_1^{sc}(q \rightarrow
qg) + A_1^{sc}(\bar{q} \rightarrow \bar{q}g) = 0$.

\section{Numerical results}
\label{capt:nu}In this section we present numerical results for the
total and the differential cross sections for $\gamma G_{KK}$
associated production at the LHC. In our numerical calculations the
running QCD coupling constant $\alpha_s(\mu)$ is evaluated at
three-loop order\cite{Amsler:2008zzb} and the CTEQ6.6M
PDFs\cite{Nadolsky:2008zw} are used throughout. Only $u$ and $d$
flavor are activated since numerical calculations show that
contributions from other flavor can be omitted. We take the LED
parameters $M_D$ and $\delta$ as input. Except for the scale
uncertainty plot, both the renormalization and the factorization
scale are fixed at $p_T^{\gamma}$, which is the transverse momentum
of the photon. Jet are defined by the following requirements:
\begin{eqnarray}
p^{jet}_T & > & 20\textrm{GeV}, \nonumber\\
|\eta^{jet}| & < & 2.5.
\end{eqnarray}
Besides, the following cuts are assumed in our
calculations\cite{Giudice:1998ck, Ball:2007zza}:
\begin{eqnarray}
&&p_{T}^{\gamma} > p_T^{min},\nonumber\\
&&|\eta| < 2.4,\nonumber\\
&&p_{T}^{miss} > p_T^{min},\nonumber\\
&&\Delta \phi(\gamma, p_T^{miss}) > 2.5.
\end{eqnarray}
Here the default value of $p_T^{min}$ is 400GeV in the following
calculations, as suggested in Ref.\cite{Giudice:1998ck,
Ball:2007zza}, $\eta$ is the pseudorapidity of the photon and
$p_{T}^{miss}$ is the missing transverse momentum, defined as
\begin{displaymath}
p^{miss}_T \equiv \left\{
\begin{array}{ll}
p^{\gamma}_T, &  \text{no jet in the final state},\\
p^{G}_T,      &  \text{with jet in the final state},
\end{array}
\right.
\end{displaymath}
where $p_T^G$ is the transverse momentum of the graviton. We also
require the photon to be isolated by requiring the separation of the
photon and the radiated parton $\Delta R \equiv \sqrt{\Delta \phi^2
+ \Delta \eta^2}$ to be greater than 0.4.

Moreover, it should be noted that the LED model is an effective low
energy theory. Therefore we present two classes of numerical results
to quantify the ultraviolet sensitivity: one with the truncation
$m^2_{\gamma G_{KK}}<M^2_D$, $m_{\gamma G_{KK}}$ being the invariant
mass of the graviton and the photon, while the other one is not
truncated. As pointed out in Ref.\cite{Giudice:1998ck} if the two
results significantly differ the contributions arising from regions
above $M_D$ dominate, and the calculations are not under control but
if they do not the LED model is viable.

\begin{figure}
\includegraphics[scale=0.75]{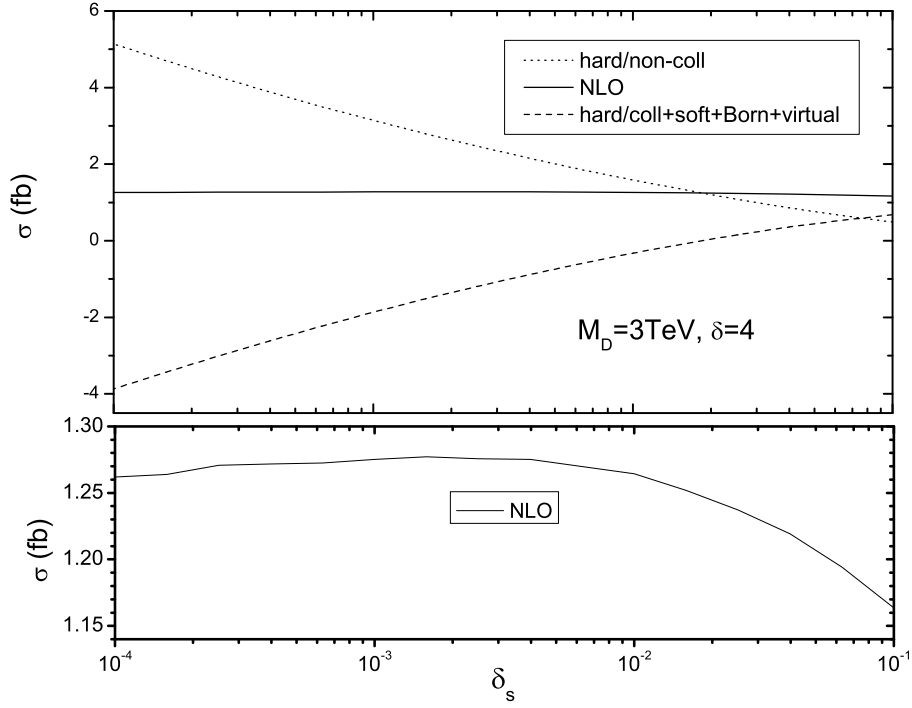}
\caption{\label{deltas}Dependence of the NLO total cross sections
for the $\gamma G_{KK}$ associated production at the LHC on the
theoretical cutoff $\delta_s$ with $\delta_c = \delta_s/50$,
assuming $M_D = 3$TeV, $\delta = 4$. Truncation $m^2_{\gamma
G_{KK}}<M^2_D$ is used here.}
\end{figure}

In Fig.\ref{deltas} we show that it is reasonable to use the two
cutoff phase space slicing method in our NLO QCD calculations; i.e.,
the dependence of the NLO QCD predictions on the arbitrary cutoffs
$\delta_s$ and $\delta_c$ is indeed very weak, as was also found in
Ref.\cite{Harris:2001sx}. While the Born cross sections and the
virtual corrections are cutoff independent, both the soft and
collinear contributions and the noncollinear contributions depend
strongly on the cutoffs. However, the cutoff dependence in the two
contributions ($\sigma^{S}_m + \sigma^{coll}_m$ and
$\sigma^{\overline{HC}}_m + \sigma^{\overline{C}}_m$) nearly cancel
each other , especially for the cutoff $\delta_s$ between $10^{-4}$
and $10^{-3}$, where the final results for $\sigma^{NLO}$ are almost
entirely independent of the cutoffs. Therefore, we will take
$\delta_s = 10^{-4}$ in the numerical calculations below. Generally
$\delta_c$ being $50-100$ times smaller than $\delta_s$ is
sufficient for accurate calculations to a few
percent\cite{Harris:2001sx}, so we take $\delta_c = \delta_s/50$ in
our calculations.

\begin{figure}
\includegraphics[scale=0.75]{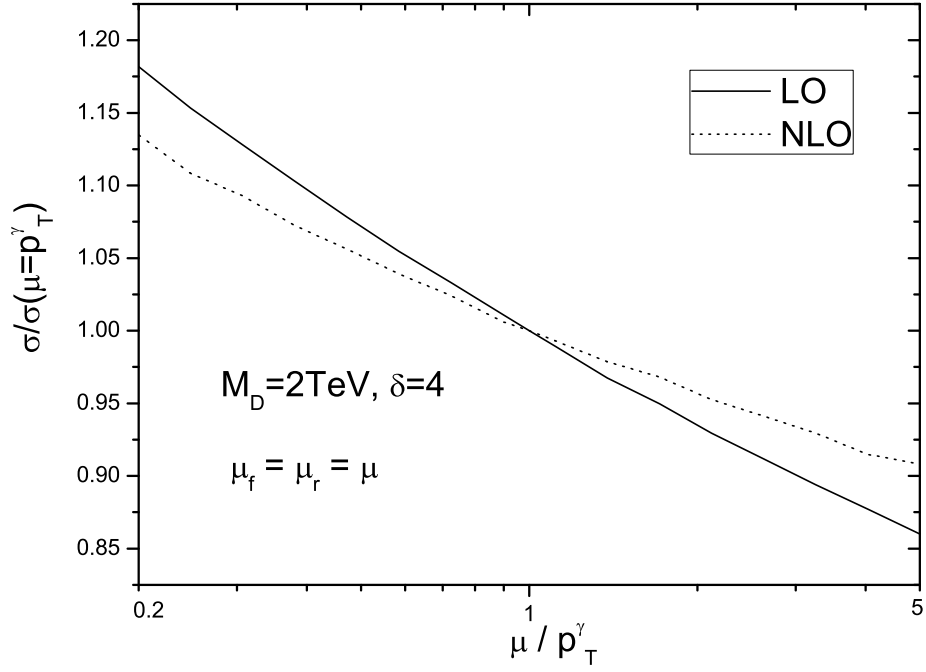}
\caption{\label{scale}Dependence of the NLO total cross sections for
the $\gamma G_{KK}$ associated production at the LHC on the
factorization scale($\mu_f$) and the renormalization scale($\mu_r$),
assuming $M_D = 2$TeV, $\delta = 4$. Truncation $m^2_{\gamma
G_{KK}}<M^2_D$ is used here.}
\end{figure}

Figure \ref{scale} shows the dependence of both the LO and the NLO
total cross sections on the factorization scale($\mu_f$) and the
renormalization scale($\mu_r$) assuming $M_D = 2$TeV, $\delta = 4$,
and setting $p_T^{min} = 400$GeV. When the scale $\mu$ varies from
$0.2p^{\gamma}_T$ to $5p^{\gamma}_T$, the LO total cross sections
vary from 1.18 to 0.86fb, while the NLO total cross sections vary
from 1.13 to 0.91fb. Thus, the NLO corrections reduce the scale
dependence, which makes the theoretical predictions somewhat more reliable.
The conclusion is similar for $\delta=2$, which is not shown
here.

\begin{figure}
\includegraphics[scale=0.9]{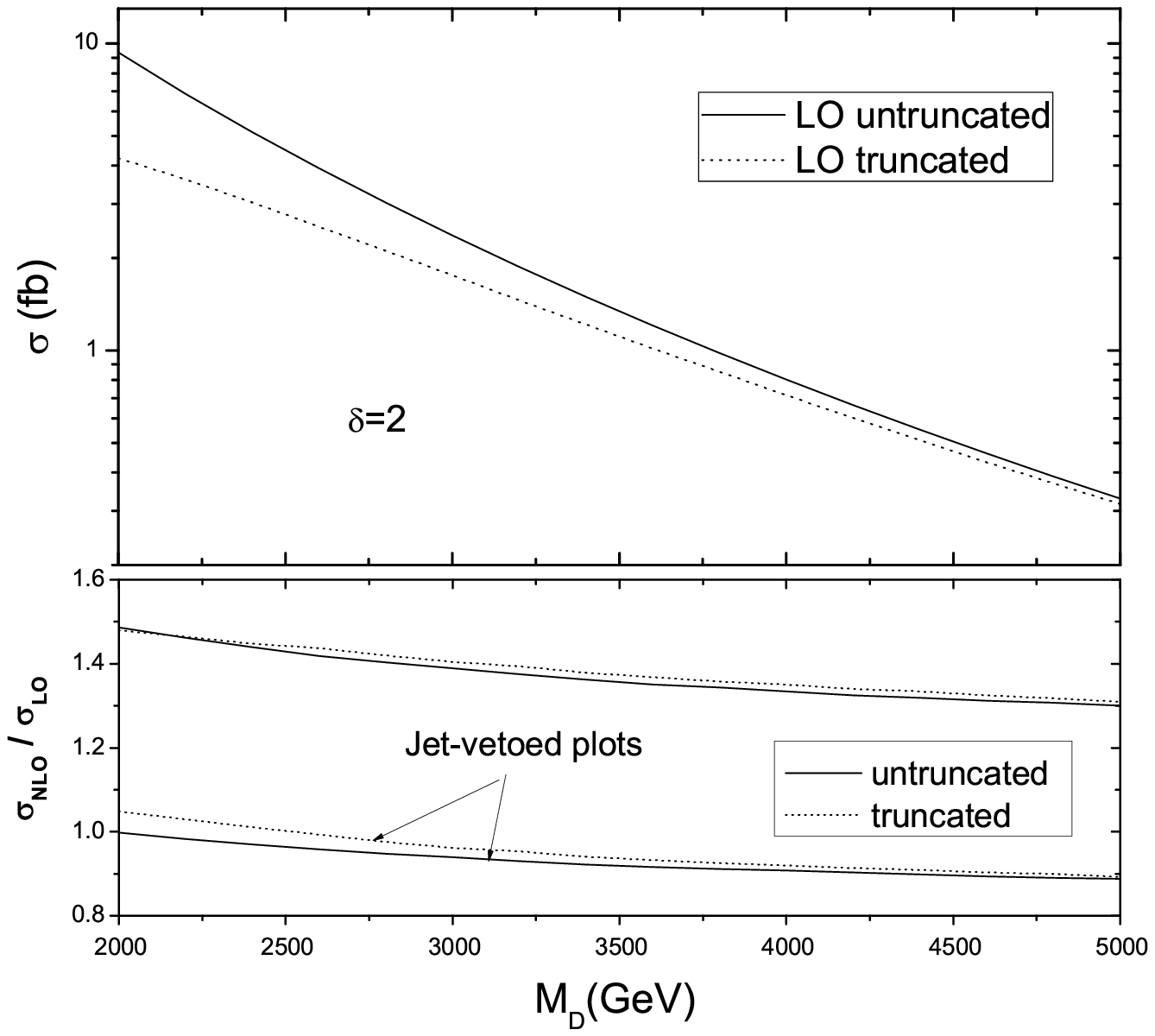}
\caption{\label{t2m}Dependence of the total cross section of the
$\gamma G_{KK}$ associated production at the LHC on $M_D$, assuming
$\delta = 2$.}
\end{figure}

\begin{figure}
\includegraphics[scale=0.9]{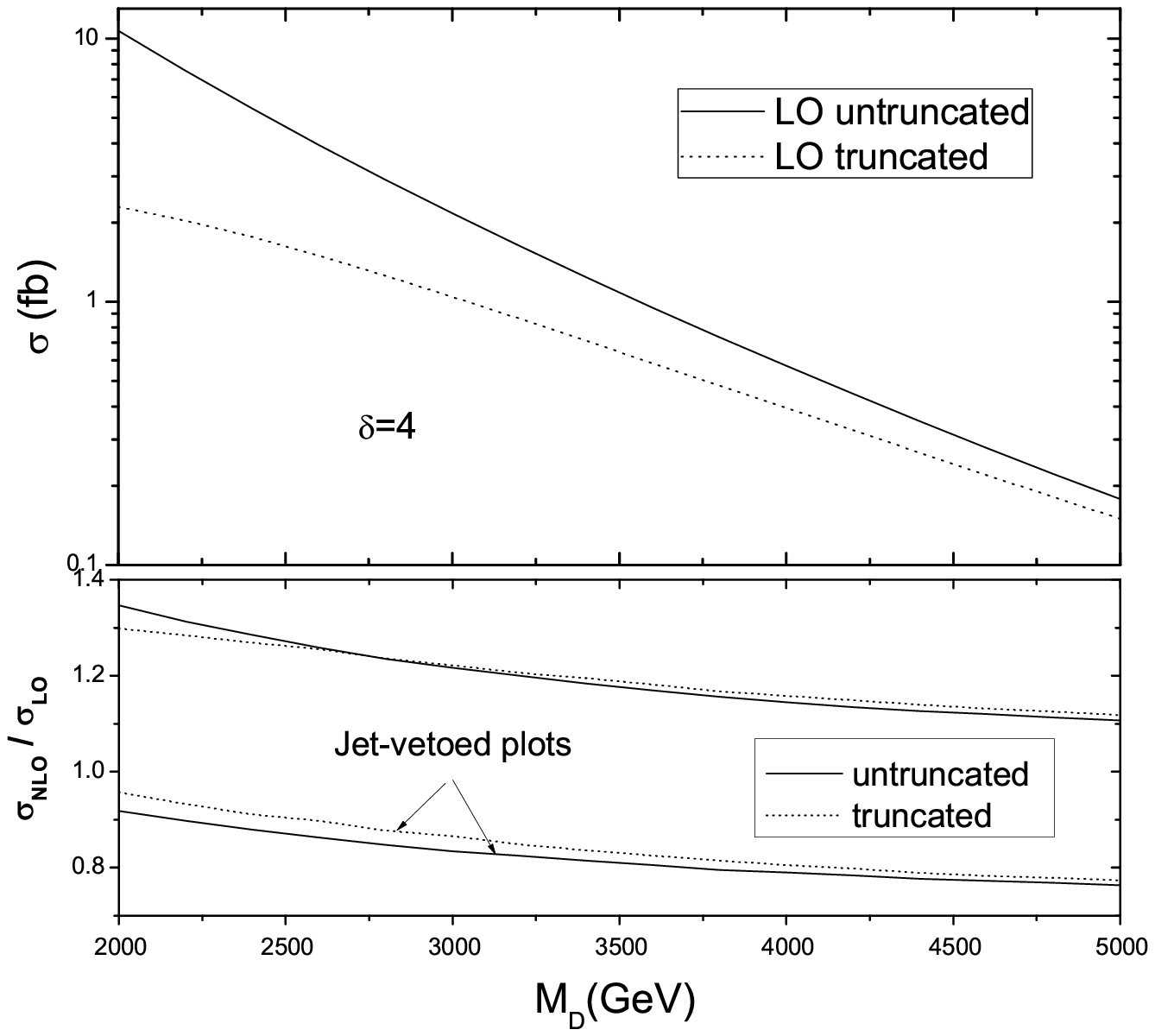}
\caption{\label{t4m}Dependence of the total cross section of the
$\gamma G_{KK}$ associated production at the LHC on $M_D$, assuming
$\delta = 4$.}
\end{figure}

In Fig.\ref{t2m} and Fig.\ref{t4m} we show the dependence of
both the LO and the NLO total cross sections on $M_D$, setting
$p^{min}_T = 400$GeV, and assuming $\delta = 2$ and 4, respectively.
As $M_D$ increases the LO total cross sections decrease and the two
results, with and without the truncation, approach each other. Also
shown in Fig.\ref{t2m} and Fig.\ref{t4m} are the $K$ factors, defined
as $\sigma^{NLO}/\sigma^{LO}$, which are around $1.3\sim1.5$ for
$\delta=2$ and $1.1\sim1.3$ for $\delta=4$, respectively. We also
give $K$ factors for cases with jet veto\cite{Ball:2007zza}, where
events with $p^{jet}_T
> 100$GeV are vetoed. In this case, the $K$ factors are around $0.9\sim1$ for
$\delta=2$ and $0.8\sim0.9$ for $\delta=4$, respectively, i.e., the
NLO corrections reduce the LO results, which is due to the fact that
the jet veto discards large positive contributions from real
emission processes.

\begin{figure}
\includegraphics[scale=0.9]{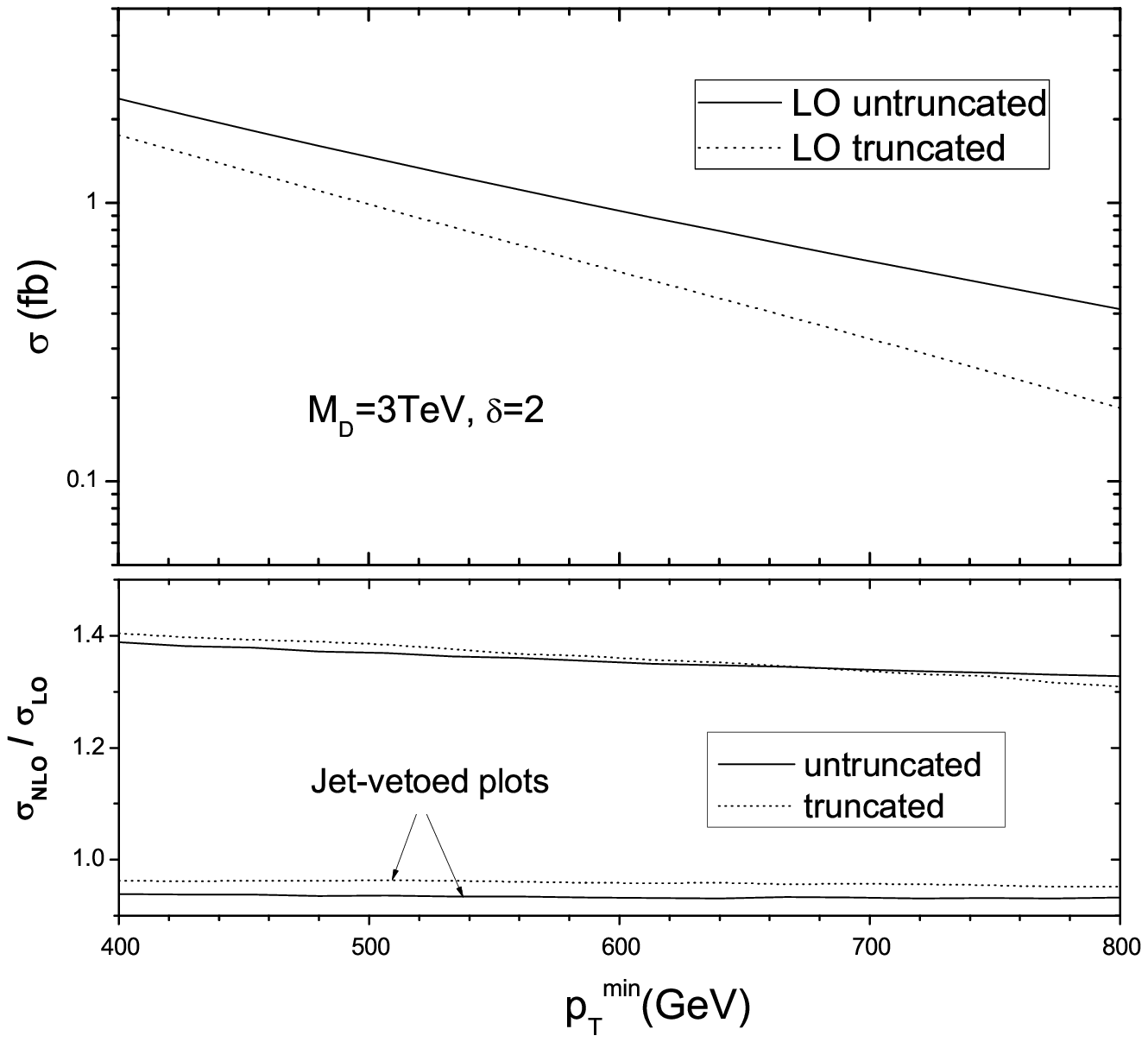}
\caption{\label{t2e}Dependence of the total cross section of the
$\gamma G_{KK}$ associated production at the LHC on $p^{min}_{T}$,
assuming $M_D = 3$TeV, $\delta = 2$.}
\end{figure}

\begin{figure}
\includegraphics[scale=0.9]{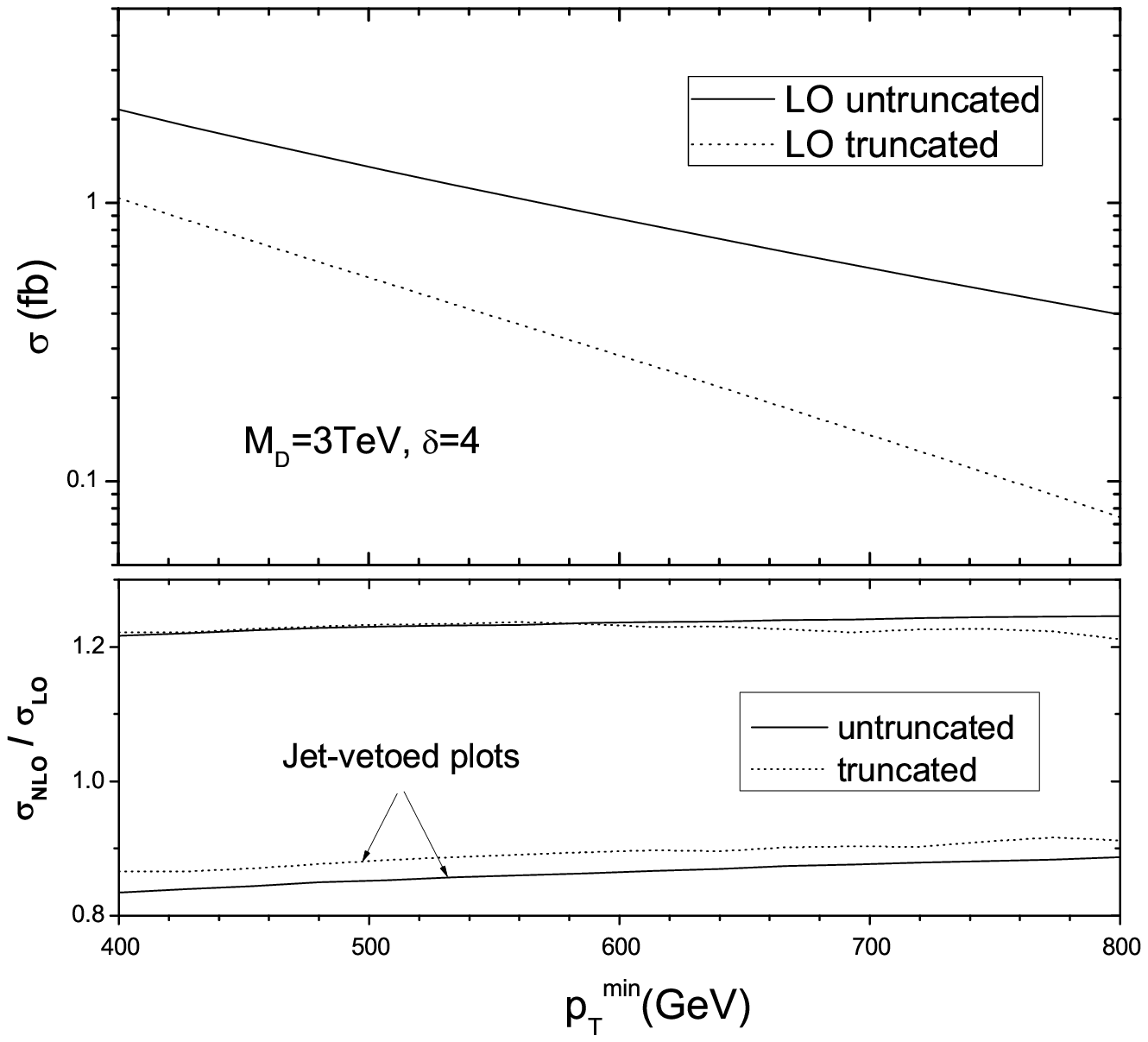}
\caption{\label{t4e}Dependence of the total cross section of the
$\gamma G_{KK}$ associated production at the LHC on $p^{min}_{T}$,
assuming $M_D = 3$TeV, $\delta = 4$.}
\end{figure}

In Fig.\ref{t2e} and Fig.\ref{t4e} we show the dependence of both
the LO and the NLO total cross sections on $p^{min}_{T}$, for
$M_D=3$TeV and $\delta = 2$ and 4, respectively. As $p^{min}_{T}$
increases the LO total cross sections decrease and the two results,
with and without the truncation, differ increasingly. The $K$ factors
are about $1.3\sim1.4$ for $\delta=2$ and 1.2 for $\delta=4$,
respectively. When jet veto is considered, the $K$ factors are around
$0.95$ for $\delta=2$ and $0.84 \sim 0.9$ for $\delta=4$,
respectively.

\begin{figure}
\includegraphics[scale=0.75]{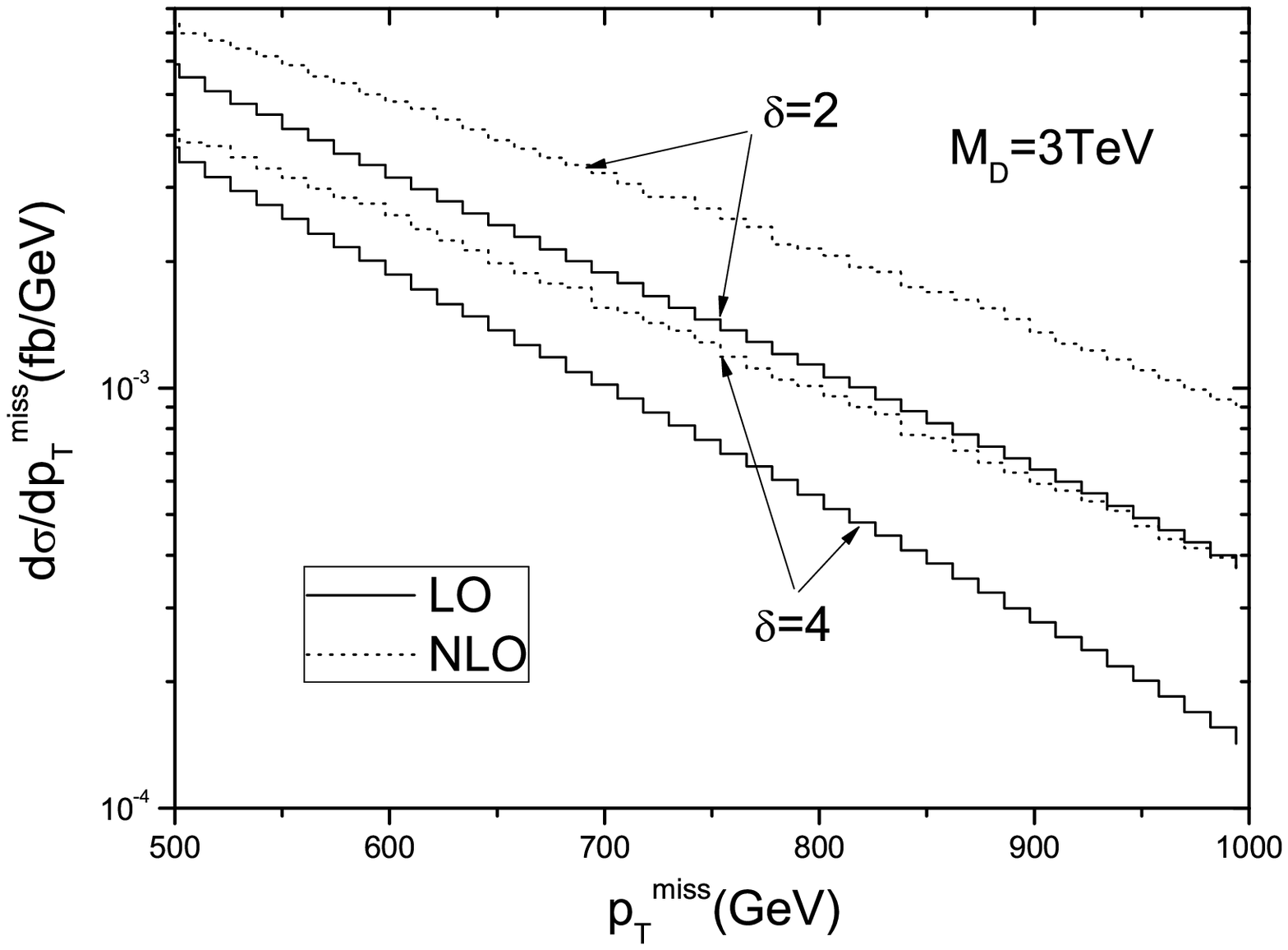}
\caption{\label{dptmiss}Dependence of the differential cross section
of the $\gamma G_{KK}$ associated production at the LHC on
$p_{T}^{miss}$, assuming $M_D = 3$TeV, $\delta=2$ and 4,
respectively.}
\end{figure}

\begin{figure}
\includegraphics[scale=0.75]{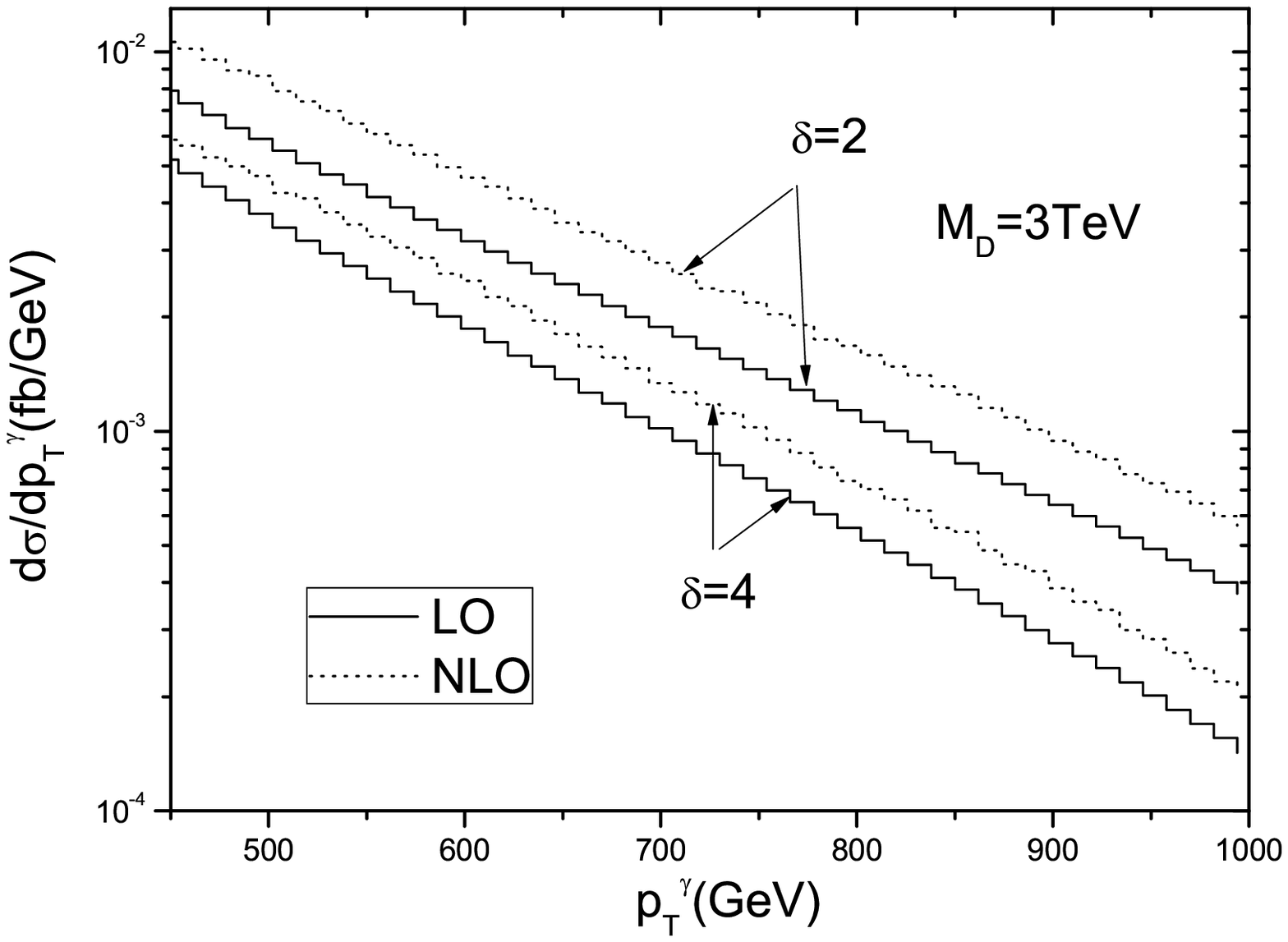}
\caption{\label{dptphoton}Dependence of the differential cross
section of the $\gamma G_{KK}$ associated production at the LHC on
$p_{T}^{\gamma}$, assuming $M_D = 3$TeV, $\delta=2$ and 4,
respectively.}
\end{figure}

\begin{figure}
\includegraphics[scale=0.75]{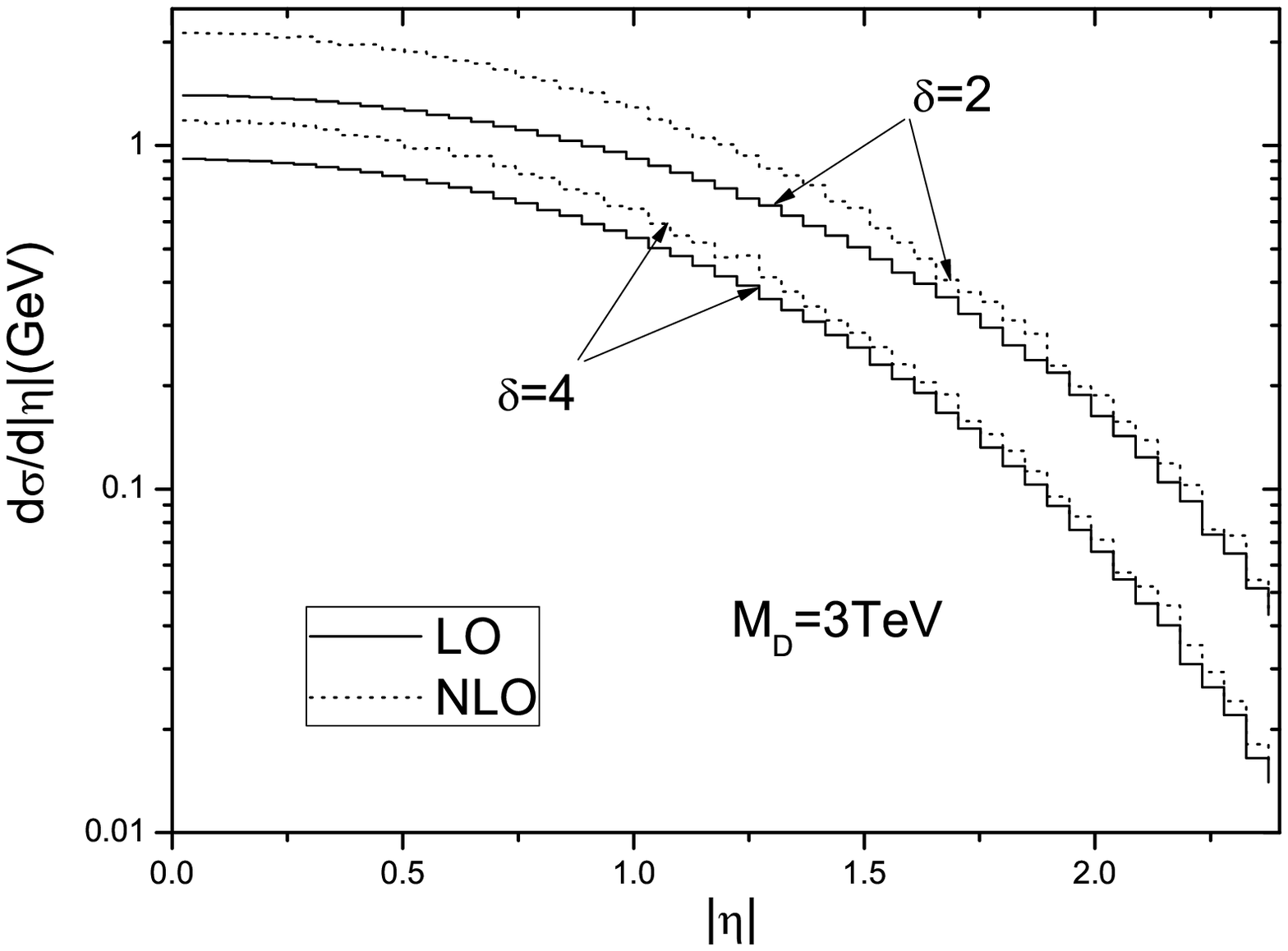}
\caption{\label{deta}Dependence of the differential cross section of
the $\gamma G_{KK}$ associated production at the LHC on $|\eta|$,
assuming $M_D = 3$TeV, $\delta=2$ and 4, respectively.}
\end{figure}

In Figs.\ref{dptmiss}-\ref{deta} we display differential cross
sections with truncation as functions of the missing transverse
momentum, the transverse momentum and the pseudorapidity of photon,
respectively. We find that the NLO QCD corrections always enhance
the LO differential cross sections but do not significantly change
the shapes of the LO differential cross sections.

In conclusion, we have calculated the complete NLO QCD corrections
to the inclusive total cross sections for $\gamma G_{KK}$ associated
production in the LED model at the LHC. The NLO corrections
generally enhance the total cross sections and the $K$ factor is
around $1.3 \sim 1.5$ for $\delta=2$ and $1.1\sim1.3$ for
$\delta=4$, respectively. When jet veto is considered, the NLO
contributions reduce the LO results, the $K$ factors are around
$0.9\sim1$ for $\delta=2$ and $0.8\sim0.9$ for $\delta=4$,
respectively. We also compared the results with and without
truncation of $m_{\gamma G_{KK}}$ to quantify the ultraviolet
sensitivity of the LED model. The NLO QCD corrections were found to
reduce the dependence of the total cross sections on the
renormalization/factorization scale. We also calculated some
important differential distributions for this process at the NLO,
including the missing transverse momentum distribution, the
transverse momentum distribution and the pseudorapidity distribution
of photon. We found that the NLO corrections enhance these LO
differential cross sections but do not appreciably change their
shapes.

\section*{Acknowledgements}
This work is supported in part by the National Natural Science
Foundation of China under grants No.10721063, No.10975004, and
No.10635030, and the US Department of Energy, Division of High
Energy Physics under Grant No. DE-FG02-91-ER4086.
\section*{APPENDIX A}
In this appendix we give the related Feynman
rules\cite{Han:1998sg}\cite{Giudice:1998ck}.
\\
\\
 ~~$\bar{q}(k_2)q(k_1)G_{\mu \nu}$: ~~~~~~$-i
\frac{\kappa}{8}[\gamma_{\mu} (k_1-k_2)_{\nu} + \gamma_{\nu}
(k_1-k_2)_{\mu}]$,
\\
\\
 ~~$V_{\alpha}(k_1)V_{\beta}(k_2)G_{\mu \nu}$: ~~$-i
\frac{\kappa}{2} [k_1 \cdot k_2 C_{\mu \nu, \alpha \beta} + D_{\mu
\nu, \alpha \beta}(k_1,k_2) + E_{\mu \nu, \alpha \beta}(k_1,k_2)]$,
\\
\\
 ~~$\bar{q}(k_2)q(k_1)V^a_{\alpha}G_{\mu \nu}$: ~~$i
\frac{\kappa}{4}g T^a (C_{\mu \nu, \alpha \beta} -
\eta_{\mu\nu}\eta_{\alpha\beta}) \gamma^{\beta}$.
\\
\\
In all the Feynman rules the particle momenta flow inward, $g T^a$ represents
either $g_s T^a$ if $V$ is a gluon or
$eQ_{f}$ if $V$ is a photon and

\begin{eqnarray}
C_{\mu \nu, \alpha \beta} & = & \eta_{\mu \alpha}\eta_{\nu \beta} +
\eta_{\mu \beta}\eta_{\nu \alpha} - \eta_{\mu \nu}\eta_{\alpha
\beta},
\nonumber\\
D_{\mu \nu, \alpha \beta}(k_1,k_2) & = & \eta_{\mu \nu} k_{1 \beta}
k_{2 \alpha} - [\eta_{\mu \beta} k_{1 \nu} k_{2 \alpha} + \eta_{\mu
\alpha} k_{1 \beta} k_{2 \nu} -\eta_{\alpha \beta} k_{1 \mu} k_{2
\nu} +(\mu \leftrightarrow \nu)],
\nonumber\\
E_{\mu \nu, \alpha \beta}(k_1,k_2) & = & \eta_{\mu
\nu}(k_{1\alpha}k_{1\beta} + k_{2\alpha}k_{2\beta} +
k_{1\alpha}k_{2\beta}) - [\eta_{\nu \beta} k_{1 \mu} k_{1 \alpha} +
\eta_{\nu \alpha} k_{2 \mu} k_{2 \beta} +(\mu \leftrightarrow \nu)].
\end{eqnarray}

\section*{APPENDIX B } We collect the explicit expressions of the finite terms of the
matrix element squared in this appendix. $\mathcal {S}$ in
Eq.(\ref{eq:virtual}) is given by

{\allowdisplaybreaks
\begin{align}
\mathcal{S} = & \frac{\kappa^2 Q_q^2 }{6} \{\frac{1}{u t}[3 C^2_0 t
(m^2-4 t) (m^4-2 t m^2+s^2+t^2) - 3 C^5_0 (m^2-4 t) (m^2-t) (m^4-2 t
m^2+s^2 \nonumber \\ & + t^2) - 3 D^1_0 s t (m^2-4 t) (m^4-2 t
m^2+s^2+t^2) + 3 D^2_0 s u (2 s^2+2 t s+t^2) (3 m^2-4 (s+t))
\nonumber \\ & + 3 C^6_0 (2 s^3+4 t s^2+3 t^2 s+t^3) (3 m^2-4 (s+t))
+ 3 C^3_0 u (2 s^2+2 t s+t^2) (4 (s+t)-3 m^2) \nonumber \\ & -3
C^4_0 (m^2-s) (m^6-6 t m^4+(-5 s^2-6 t s+6 t^2) m^2+4 s (2 s^2+3 t
s+3 t^2)) \nonumber \\ & - 3 C^1_0 ((s+8 t) m^6-6 t (s+4 t) m^4+(7
s^3+18 t s^2+30 t^2 s+32 t^3) m^2-4 (2 s^4+5 t s^3 \nonumber \\ & +9
t^2 s^2+8 t^3 s+4 t^4))] + \frac{1}{s t u}[2 ((3 s+14 t) m^6-6
(s^2+5 t s+t^2) m^4+(3 s^3+30 t s^2 \nonumber \\ & -16 t^3) m^2+2 t
(-7 s^3-3 t s^2+8 t^2 s+4 t^3)) + 18 (-(s+5 t) m^6+(s^2+8 t s+11
t^2) m^4 \nonumber \\ & -(s^3+8 t s^2+15 t^2 s+12 t^3) m^2+t (5
s^3+11 t s^2+12 t^2 s+6 t^3))] + \nonumber \\ & \frac{1}{s t u
(m^2-s)^2 (m^2-t)^2 (s+t)^2 }[3 s ((3 s^2+4 t s-2 t^2) m^{12}-2 (3
s^3+16 t s^2+17 t^2 s \nonumber \\ & +4 t^3) m^{10}+(6 s^4+22 t
s^3+56 t^2
s^2+68 t^3 s+34 t^4) m^8-2 (3 s^5-5 t s^4-41 t^2 s^3 \nonumber \\
& -46 t^3 s^2+5 t^4 s+18 t^5) m^6+(3 s^6+2 t s^5-110 t^2 s^4-300 t^3
s^3-266 t^4 s^2-60 t^5 s \nonumber \\ & +12 t^6) m^4+2 s t (-3
s^5+14 t s^4+86 t^2 s^3+137 t^3 s^2+84 t^4 s+16 t^5) m^2 \nonumber
\\ & -24 s^2 t^3 (s+t)^3) \log \left(\frac{m^2}{\mu ^2}\right) m^2-3 s (m^2-t)^2
(s+t)^2 (3 m^{10}-6 (s+3 t) m^8+6 (s^2+3 t s \nonumber \\ & +3 t^2)
m^6-2 s^2 (3 s+11 t) m^4+s^2 (3 s^2+42 t s+22 t^2) m^2-20 s^3 t
(s+t)) \log \left(\frac{s}{\mu ^2}\right) \nonumber \\ & -(m^2-s)
((m^2-t) ((s+t)^2 (21 s+34 t) m^{10}-(18 s^4+196 t s^3+483 t^2
s^2+441 t^3 s \nonumber \\ & +136 t^4) m^8+(s+t)^2 (15 s^3+165 t
s^2+466 t^2 s+238 t^3) m^6-(18 s^6+124 t s^5+587 t^2 s^4 \nonumber
\\ & +1473 t^3 s^3+1786 t^4 s^2+998 t^5 s+204 t^6) m^4+t (s+t)^2 (49
s^4+154 t s^3+382 t^2 s^2 \nonumber \\ & +340 t^3 s+68 t^4) m^2-2 s
t^2 (s+t)^3 (11 s^2+34 t s+34 t^2)-3 (m^2-s) s (m^2-t) t ((2 s
\nonumber
\\ & +5 t) m^6-6 t (s+t) m^4+3 (2 s-t) (s+t)^2 m^2-4 (2 s-t) (s+t)^3)
\log \left(-\frac{u}{\mu ^2}\right)) \nonumber \\ & -3 s t (s+t)^2
(3 (4 s+3 t) m^8-(24 s^2+57 t s+22 t^2) m^6+(12 s^3+75 t s^2+74 t^2
s \nonumber
\\ & +17 t^3) m^4-t (27 s^3+64 t s^2+33 t^2 s+4 t^3) m^2+4 s t^2 (3
s^2+4 t s+t^2)) \log \left(-\frac{t}{\mu ^2}\right))]\},
\end{align}
}
where $Q_q$ is the electric charge of the initial (anti)quark, and
\begin{eqnarray}
C^1_0 & = & -\frac{\pi^2}{3s},
\nonumber\\
C^2_0 & = & \frac{1}{2t}(\log^2 \left(\frac{-t}{s}\right) + \frac{\pi^2}{3}),
\nonumber\\
C^3_0 & = & \frac{1}{2u}(\log^2 \left(\frac{-u}{s}\right) + \frac{\pi^2}{3}),
\nonumber\\
C^4_0 & = & \frac{\log^2 \left(\frac{m^2}{s}\right)}{2(m^2-s)},
\nonumber\\
C^5_0 & = & \frac{1}{2(m^2-t)}[\log \left(\frac{m^2}{-t}\right)(\log
\left(\frac{m^2}{s}\right) + \log \left(\frac{-t}{s}\right)) -
\pi^2],
\nonumber\\
C^6_0 & = & \frac{1}{2(m^2-u)}[\log \left(\frac{m^2}{-u}\right)(\log
\left(\frac{m^2}{s}\right) + \log \left(\frac{-u}{s}\right)) -
\pi^2],
\nonumber\\
D^1_0 & = & \frac{1}{st}[- \log^2 \left(\frac{m^2}{s}\right) + \pi^2
-2\text{Li}_2(1-\frac{m^2}{s}) -2\text{Li}_2(1-\frac{m^2}{t})],
\nonumber\\
D^2_0 & = & \frac{1}{su}[- \log^2 \left(\frac{m^2}{s}\right) + \pi^2
-2\text{Li}_2(1-\frac{m^2}{s}) -2\text{Li}_2(1-\frac{m^2}{u})].
\end{eqnarray}

\bibliography{GammaG}

\end{document}